**Casimir force between two dielectric layers: Van Kampen approach**

M.V. Davidovich

Saratov National Research State University named after N.G. Chernyshevsky

E-mail: DavidovichMV@info.sgu.ru

The Van Kampen method is used to calculate the Casimir force for two dielectric layers. Several terms of Lorentz oscillators are used in the permittivity model. A conductive dielectric (metal) with the Drude model is considered as a special case. The dependence of strength on thickness has a complex character with saturation at thicknesses of the order of 10 nm. At low thickness, the force density is proportional to the square of the thickness, but this is the case at low thicknesses, when the continuum model is no longer applicable. The correspondence between the method of the Casimir model and the Lorentz model is shown, as well as its applicability for an arbitrary configuration of layers and for a finite temperature.

Two parallel dielectric layers with thicknesses $t_1$ and $t_2$ at a distance $d$ between them and at zero temperature in a large resonator with ideal walls Fig. 1 can interact with the Casimir force, determined similarly to [1,2], namely by summing the perturbed natural eigenfrequencies of the resonator ($\alpha, \beta = e, h$) and determining the energy of zero oscillations

$$\tilde{E}(d) = \frac{\hbar}{2} \mathrm{Re} \sum_{\substack{\alpha=e,h \\ \beta=e,h}} \sum_{mnl} \tilde{\omega}_{mnl}^{\alpha\beta}(d). \qquad (1)$$

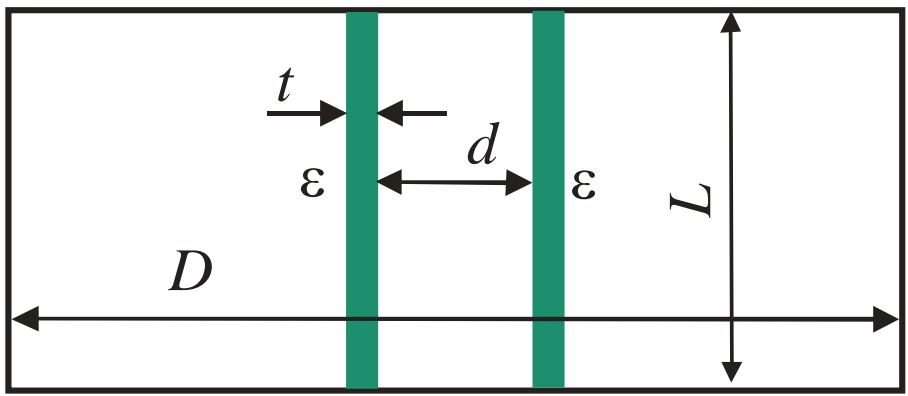

Fig. 1. Rectangular resonator with two dielectric layers at $L_x = L_y = L$ and $L_z = D$

Dependency $\exp(i\omega t)$ is used. Formula (1) was used by Casimir for ideally conducting plates, while radiated modes in the form of standing waves were considered, which can only be in such a structure (in an ideal resonator). At a finite temperature, taking into account the average energy of the quantum oscillator [3] $\theta\!\left(\tilde{\omega}_{mnl}^{\alpha\beta},T\right)=\left(\hbar\tilde{\omega}_{mnl}^{\alpha\beta}/2\right)\coth\!\left(\hbar\tilde{\omega}_{mnl}^{\alpha\beta}/(2k_B T)\right)$, we should take

$$\tilde{E}(d)=\mathrm{Re}\sum_{\substack{\alpha=e,h\\\beta=e,h}}\sum_{mnl}\theta\!\left(\tilde{\omega}_{mnl}^{\alpha\beta},T\right)\tilde{\omega}_{mnl}^{\alpha\beta}(d).$$

Further, it will be seen that when summing in this case by Van Kampen, the correct result is obtained in the form of a sum over the Matsubara frequencies. Taking the real part in expressions like (1) is based on the fact that there is a stationary equilibrium energy density of the resonator $\tilde{E}(\omega)=\varepsilon_0\varepsilon'(\omega)|\mathbf{E}(\omega)|^2/2$ if there is no accumulated kinetic energy of charges moving under the action of the field [4–9]. After quantization of the field, the energy can be represented as the sum of the oscillator energies (1). This is not true for plasmas [4,9], but for collisional plasmas, the formula for the density of the stored average energy over a period (including kinetic energy) has the form [9]

$$\tilde{E}(\omega)=\frac{\varepsilon_0|\mathbf{E}|^2}{4}\left(1+\frac{\omega_p^2}{\omega^2+\omega_c^2}+\sqrt{\left(1-\frac{\omega_p^2}{\omega^2+\omega_c^2}\right)^2+\frac{\omega_p^4\omega_c^2}{\left(\omega^2+\omega_c^2\right)^2\omega^2}}\right),$$

which can also be represented as a set of oscillator energies (1), and at a low collision frequency (CF) $\omega_c$ we get $\tilde{E}(\omega)=\varepsilon_0|\mathbf{E}(\omega)|^2/2$. A similar formula holds for Lorentz oscillators [4]. If the CF tends to zero, the result (1) is always real. When the resonator is expanded to free space, we obtain the interaction force of the plates in it. Along the plates there are plasmon-polaritons (PP) that contribute to the force. They are slow (in the form of attenuating or evanescent modes) and fast (in the form of radiated or leakage modes). At short distances, the main contribution are made by slow PPs, and at large ones by fast radiated PPs. For small thicknesses, it should be expected that the force density is proportional to the product of the thicknesses $t_1 t_2$. Further, to simplify the formulas, we consider the plates to be the same. For the force density, we should take $\partial_d \tilde{E}(d)/\left(L_x L_y\right)$, where $L_x L_y$ is the large area of the layers. The dielectric material is considered to be the same and has a spectral dielectric permittivity (DP) $\varepsilon(\omega)=\varepsilon'(\omega)-i\varepsilon''(\omega)$, moreover $\varepsilon''(0)=0$, if there are no free charge carriers (for plasma $\varepsilon''(0)=\infty$). The tilde indicates the frequencies, energy, and longitudinal wavenumber perturbed by the dielectric. In an empty resonator with dimensions $L_x$, $L_y$, $L_z$, there are undisturbed resonant frequencies $\omega_{mnl}^{e,h} = c\sqrt{k_{xn}^2 + k_{yn}^2 + k_{zl}^2}$ of TE$_{mnl}$ (or H$_{mnl}$) modes, where $k_{xn} = m\pi/L_x$, $k_{yn} = n\pi/L_y$, $k_{zl} = l\pi/L_z$, $m$=0.1,..., $n$=0.1,..., $l$=1.2,..., except for $m$=$n$=0, as well as frequencies $\omega_{mnl}^e = c\sqrt{k_{xn}^2 + k_{yn}^2 + k_{zl}^2}$ of TM$_{mnl}$ (or E$_{mnl}$) modes, the difference is which is that now $m$=1,2,..., $n$=1,2,..., $l$=0,1,2,.... [4–6]. Thus, oscillation degeneracy takes place in an empty resonator. In a filled resonator, it is removed: $\widetilde{k}_{zl} = k_{zl} + \Delta k_{zl}$ is the value perturbed by the dielectric, $\Delta k_{zl} \sim 1/L_z$. Going to the limit $L_{x,y} \to \infty$ means continuity of the transverse indices $dk_{xm} = dk_x = (\pi/L_x)dm$, $dk_{yn} = dk_y = (\pi/L_y)dn$, and replacing the two-dimensional sum in (1) with a two-dimensional integral. It is convenient to switch to the polar coordinates $k_x = \kappa \cos(\varphi)$, $k_y = \kappa \sin(\varphi)$. Then the angle integral is calculated and is equal to $2\pi$. The transition to the limit $L_z \to \infty$ reduces the sum of the expansion of the resonator to the entire space to a two-dimensional integral over $dk_z d\kappa = (\pi/L_z)dl$, $dk_z = dk_{zl} = (\pi/L_z)dl$. The frequencies in the finite resonator are discrete. They lie in the upper half-plane of the complex frequency plane symmetrically relative to the imaginary axis $\pm\omega_n' + i\omega_n''$ [10]. Any real damped oscillation can be represented as the sum of the oscillations with these frequencies, for example $\cos(\omega_n' t)\exp(-\omega_n'' t)$. The radiation corresponds to a change in the sign of the imaginary part. These frequencies lie in the lower half-plane. In a resonator with infinite walls and with a finite size $L_z$, the characteristic equations define frequencies $\widetilde{\omega}_l = \widetilde{\omega}_l(\kappa)$ as continuous meromorphic functions, $l$=1,2,.... For $L_z \to \infty$ and plates in free space the characteristic equations $f_{e,h}(\kappa, \widetilde{k}_z) = 0$, which are functions of two variables, and are the dispersion equations (DE) of plasmon-polaritons (PP). In a vacuum $\kappa^2 = k_0^2 - k_z^2$, and in a dielectric $\kappa^2 = k_0^2\varepsilon - k_z^2$. The value $\widetilde{k}_z$ of the structure is determined from such DE. Considering the dissipation to be extremely small, it can be real $\widetilde{k}_z < k_0$ (fast leakage plasmon-polaritons, or modes emitted in vacuum), and imaginary, which determines slow surface plasmon-polaritons along the surface. The frequencies perturbed by the dielectric are defined as $\widetilde{\omega} = c\sqrt{\kappa^2 + \widetilde{k}_z^2} = ck_0\sqrt{1 + (2k_z + \Delta k_z^2)/k_0^2}$. Then we can consider DE as a function of $\kappa$ and $k_0$: $f_{e,h}(\kappa, k_0)$, $k_0 = \omega/c$. The DE for the E-mod (TM-mod) and H-mod (TE-mod) are obtained by mode matching and have the form [11–13]

$$f_h(\kappa, k_0, d, t) = \exp(2K_0 d)\left[\frac{2K_0 K \coth(Kt) + K^2 + K_0^2}{K^2 - K_0^2}\right]^2 - 1, \qquad (2)$$

$$f_e(\kappa, k_0, d, t) = \exp(2K_0 d)\left[\frac{2\varepsilon K_0 K \coth(Kt) + K^2 + \varepsilon^2 K_0^2}{K^2 - \varepsilon^2 K_0^2}\right]^2 - 1. \qquad (3)$$

Here $K_0 = \sqrt{\kappa^2 - k_0^2}$, $K = \sqrt{\kappa^2 - k_0^2 \varepsilon(\omega)}$. it is convenient to denote $k = ik_0$ and move on to the complex frequency $\xi = i\omega$. Then $K_0 = \sqrt{\kappa^2 + k^2}$, $K = \sqrt{\kappa^2 + k^2 \varepsilon(-ick)}$. Note that there are several possible forms of the characteristic equation, which are also the DE for plasmon-polaritons [14]. The force density or Casimir pressure is defined as $P(d) = -\partial_d \tilde{E}(d)$. According to the method [11] (see also [12,13]), we have the equation (in [12,13], the multiplier 2 was lost, although it was later restored)

$$P(d) = -\frac{\hbar c}{2\pi^2} \int_0^\infty \kappa d\kappa \int_0^\infty K_0 \left( \frac{1}{f_e(\kappa, k)} + \frac{1}{f_h(\kappa, k)} \right) dk . \qquad (4)$$

In its meaning, it determines the pressure between the layers. Negative pressure means attraction. Equality (4) is obtained from the principle of argument (argument theorem) in the form

$$\left[ \sum_n \tilde{\omega}_n(\kappa) \right]_{f_{e,h}(\tilde{\omega}_n) = 0} = \frac{1}{2\pi i} \oint_C \frac{f'_{e,h}(\omega)}{f_{e,h}(\omega)} \omega d\omega = \frac{-1}{2\pi} \oint_C [\ln(f_e(\omega)) + \ln(f_h(\omega))] d(i\omega) =$$
$$= \frac{c}{2\pi} \int_{-\infty}^\infty [\ln(f_e(\kappa, k, d)) + \ln(f_h(\kappa, k, d))] dk = \frac{c}{\pi} \int_0^\infty [\ln(f_e(\kappa, k, d)) + \ln(f_h(\kappa, k, d))] dk \qquad (5)$$

By taking

$$\tilde{E}(d) = \frac{\hbar}{2} \frac{L_x L_y}{2\pi} \mathrm{Re} \int_0^\infty \sum_n \left( \tilde{\omega}_n^e(\kappa, d) + \tilde{\omega}_n^h(\kappa, d) \right) \kappa d\kappa ,$$

and $P(d) = -\partial_d \tilde{E}(d) / (L_x L_y)$, we get (4). In this case, it should be taken into account that the contour is chosen along the imaginary axis and the right semicircle with a counterclockwise bypass, as well as the fact that the zeros $f_{e,h}(\kappa, k, d) = 0$ (for given $\kappa$) are located symmetrically relative to the imaginary axis in the upper half-plane (in the right half-plane $\xi$ they are complex conjugate). Therefore, the sum is always real, and it is not necessary to take the real part (due to the parity of the equations, there are also roots of the opposite sign). The contour in the plane $\xi$ can be drawn as in Fig. 3.7 of [15], while the real parts of the frequencies are taken into account twice. When the average energy $\theta(\omega)$ is taken into account, the additional poles appear on the imaginary axis $\xi$. As a result of which the force density is calculated using the sum of the Matsubara frequencies. However, the result (4) is determined up to the multipliers $A_{e,h}(\kappa, k)$, because for any multiplier $A_{e,h}^{-1}(\kappa, k) f_{e,h}(\kappa, k, d) = 0$. The multipliers should be determined from the condition that in the limiting case the Casimir problem [1] or Lifshitz problem [16] is obtained. So, for $\varepsilon \to \infty$ from (2), (3) we obtain the Casimir problem $f_e(\kappa, k, d) = f_h(\kappa, k, d) = \exp(2K_0 d) - 1$, $P(d) = -\hbar c \pi^2 / (240 d^4)$, i.e. in this case $A_{e,h}(\kappa, k) = 1$. In the absence of plates ($t=0$ or $\varepsilon=1$) we have $f_{e,h}(\kappa,k,d)\to\infty$ and $P(d)=0$. With a small plate thickness, we have

$$f_h^{-1}(\kappa,k_0)\approx t^2\left(K^2-K_0^2\right)^2/\left[4K_0^2\exp(2K_0d)\right],$$

$$f_e^{-1}(\kappa,k_0)=t^2\left(K^2-\varepsilon^2K_0^2\right)/\left[4\varepsilon^2K_0^2\exp(2K_0d)\right],$$

and the force is proportional to the square of the thickness. The Considering plates of different thicknesses leads to a proportional force to $t_1t_2$. Note that for the Lifshitz problem [16] on the gap between dielectric half-planes $t\to\infty$, $\coth(Kt)=1$, and we obtain

$$f_h(\kappa,k_0)=\exp(2K_0d)\left(\frac{K+K_0}{K-K_0}\right)^2-1, \qquad (6)$$

$$f_e(\kappa,k_0)=\exp(2K_0d)\left(\frac{K+\varepsilon K_0}{K-\varepsilon K_0}\right)^2-1. \qquad (7)$$

For Casimir problem $\varepsilon\to\infty$, $K\to\infty$, и $f_e(\kappa,k_0)=f_h(\kappa,k_0)=\exp(2K_0d)-1$. The relations (6), (7) can be represented in terms of the reflection coefficients of the modes [15].

For further development, the DP model should be used. The dispersion of real dielectrics over a wide range is usually quite complex. Using Lorentz's law of dispersion, taking into account the internal field, it can be represented as the Clausius-Mossotti formula.

$$\varepsilon(\omega)=\frac{1+\dfrac{2}{3}\sum_{n,m=1}\dfrac{\omega_{pm}^2}{\omega_{mn}^2-\omega^2+i\omega_{cmn}\omega}}{1-\dfrac{1}{3}\sum_{n,m=1}\dfrac{\omega_{pm}^2}{\omega_{mn}^2-\omega^2+i\omega_{cmn}\omega}}. \qquad (8)$$

Here we used the Lorentz polarizability for an atom with transition frequencies $\omega_{mn}$

$$\alpha_{mn}=\frac{e^2N_m}{\varepsilon_0m_e}\frac{1}{\omega_{mn}^2-\omega^2+i\omega\omega_{cmn}}$$

and the Lorentz-Lorentz formula for the internal field. The introduction of an internal field implies the absence of resonances, which is not performed in a wide range. The frequencies $\omega_{cmn}$ characterize the relaxation times of the levels. If the concentrations of atoms $N_m$ of the $m$ variety (or squares $\omega_{pm}^2$ of plasma frequencies (PF)) are small, i.e. the sum is small compared to unity, (8) can be decomposed into a small parameter:

$$\varepsilon(\omega)\approx 1+\sum_{n,m=1}\frac{\omega_{pm}^2}{\omega_{mn}^2-\omega^2+i\omega_{cmn}\omega}. \qquad (9)$$

This formula is derived from the Lorentz oscillator model [4] and is often used, although it is strictly valid for a rarefied gas of oscillators with several resonant frequencies $\omega_{mn}$. Next, we use it, since formula (8) leads to inadequate results at resonances (small distances). It can be used in the low-frequency range at high $d$. If there are atoms of only one kind, then $m = 1$. If there is only one resonant frequency, then $n = 1$. The values $\omega_{pm}^2$ characterize the oscillator forces calculated from solving a quantum mechanics problem. If there is $\omega = \omega_{mn}$ and for small CF $\varepsilon(\omega_{mn}) \approx -2 < 0$. Formula (8) cannot be used in this case, as in the case of equality of the sum to three ($\varepsilon = \infty$), since it is obtained in the approximation of a small sum. In real media, with a large number of frequencies, significant losses, and small oscillator forces for most oscillations, the real part $\varepsilon'$ of the DP does not go through zero. Such a transition usually takes place in metals. Consideration of media with variance (9) is of interest [17]. Note that for the region significantly lower than the resonant frequencies, the optical part of the DP is obtained, determined by the polarization of the substance:

$$\varepsilon_L = 1 + \sum_{n,m=1} \frac{\omega_{pm}^2}{\omega_{mn}^2}. \tag{10}$$

The squares of PF $\omega_{pm}^2$ determine the concentrations of atoms. For metals, there are free electrons. In the model, this means a zero resonant frequency (no coupling), which characterizes additional electronic susceptibility

$$\chi_e = -\frac{\omega_p^2}{\omega^2 - i\omega_c \omega}, \tag{11}$$

determined by the PF and CF for conduction electrons of conductivity. For them, the resonant frequency is zero because they are free and not bound to atoms. Note that from (9) it is also possible to obtain the Debye dispersion law in the limit for absolutely rigid dipoles (high transition frequencies) with orientational polarization [7]. The considered models allow us to accurately describe the real media, if we take into account a sufficient number of members. Actually condensing atomic spectra have many (infinitely many) terms. Additional spectral terms arise for polyatomic systems and molecules, so it is easier to determine DP through the absorption spectrum [16, 18], which can be experimentally measured in a wide range. However, this is inconvenient for analytical and numerical calculations. Taking into account a sufficient number of terms allows us to build an adequate model of the dispersion forces. The transition to complex frequency means dependence

$$\varepsilon(k) = 1 + \chi_e(k) + \sum_{n,m=1} \frac{k_{pm}^2}{k_{mn}^2 + k^2 + k_{cmn}k}, \tag{12}$$

where the corresponding wave numbers are entered. Also $\chi_e(k) = k_p^2 / (k^2 + k_c k)$. This value has poles at $k = -k_c$ and at $k = 0$. To avoid the latter, the Drude-Smith model can be used [19,20].

In finite structures, a free electron cannot escape to infinity from an atom, i.e. it can be approximately characterized by a very small coupling constant $k_s^2$ related to size, and susceptibility $\chi_e(k) = k_p^2 / (k_s^2 + k^2 - k_c k)$ can be introduced. You can take it $k_s \sim k_c$, but with a very large thickness $k_s \sim 1/t$. It's important that $\varepsilon(\infty) = 1$. This means that for $k \to \infty$ we have $\varepsilon \to 1$, and in formulas (2), (3) $K \to K_0$ and $f_{e,h}(\kappa, k) \to \infty$, providing, along with a large factor $\exp(2K_0 d)$, the convergence of the integral (4). Other DP models are possible, including accounting for the internal field, for example, according to the Onsager formula [4].

To numerically calculate the integral (4), we turn to the polar coordinates $\kappa = \chi \cos(\theta)$, $k = \chi \sin(\theta)$, $K_0 = \chi$. At the point $\theta = 0$ we have $k = 0$ and $\varepsilon(0)$ is the low-frequency DP value. At $\theta \to 0$ the DP's commitment to $\varepsilon(0)$ provides a significant contribution to strength. At all other points $\theta > 0$ the DP tends to unity at $\chi \to \infty$. Therefore, the angle integral is divided into two intervals $(0, \theta_0)$ and $(\theta_0, \pi/2)$. In the first case, we perform careful integration by angle, and $K = \chi \sqrt{1 + \sin^2(\theta)\varepsilon(\chi, \theta)} \approx \chi \sqrt{1 + \varepsilon(0)}$ if the angle $\theta_0$ is small. The ratio (2), (3) for large numbers $\chi$ is written as $f_{e,h}(\chi, \theta) = \exp(2\chi d)\varphi_{e,h}(\chi, \theta) - 1 \approx \exp(2\chi d)\varphi_{e,h}(\chi, 0)$. We select the integration areas $0 < \chi < \chi_0$ and $\chi_0 < \chi < \infty$. For the second region, considering $\chi_0$ to be a large value, we have independent from $\chi$ functions

$$\varphi_h(\chi, 0) = \left[ \frac{2\sqrt{1 + \varepsilon(0)} + 2 + \varepsilon(0)}{\varepsilon(0)} \right]^2, \tag{13}$$

$$\varphi_e(\chi, 0) = \left[ \frac{1 + 2\varepsilon(0)\sqrt{1 + \varepsilon(0)} + \varepsilon(0) + \varepsilon^2(0)}{1 + \varepsilon(0) - \varepsilon^2(0)} \right]^2, \tag{14}$$

and the result for the integral of the remainder is

$$\int_{\chi_0}^{\infty} \chi^2 \exp(-2\chi d) \left( \frac{1}{\varphi_e(\chi, 0)} + \frac{1}{\varphi_h(\chi, 0)} \right) d\chi =$$
$$= \exp(-2d\chi_0) \left( \frac{1}{\varphi_e(\chi_0, 0)} + \frac{1}{\varphi_h(\chi_0, 0)} \right) \left( \frac{\chi_0^2}{2d} + \frac{\chi_0}{2d^2} + \frac{1}{4d^3} \right).$$

This result allows us to choose $\chi_0$ so that the integral of the remainder is significantly less than the integral in the domain $0 < \chi < \chi_0$. For an area $(\theta_0, \pi/2)$, it is enough to take several points of integration along the angle. 500 were used in the calculations. The integral can even be calculated approximately by the mean value theorem at a point $\tilde{\theta} = (\theta_0 + \pi/2)/2$. Then $K = \tilde{K}(\chi) = \chi \sqrt{1 + \varepsilon(\chi, \tilde{\theta})\sin^2(\tilde{\theta})}$, and for the integral over $\chi$ we have

$$\int_0^\infty \left( \frac{1}{\exp(2\chi d)\varphi_e(\chi,\tilde{\theta})-1} + \frac{1}{\exp(2\chi d)\varphi_h(\chi,\tilde{\theta})-1} \right)\chi^2 d\chi =$$

$$\int_0^{\chi_0} \left( \frac{1}{\exp(2\chi d)\varphi_e(\chi,\tilde{\theta})-1} + \frac{1}{\exp(2\chi d)\varphi_h(\chi,\tilde{\theta})-1} \right)\chi^2 d\chi + \quad . \qquad (15)$$

$$+ \exp(-2d\chi_0)\left( \frac{1}{\varphi_e(\chi_0,\tilde{\theta})} + \frac{1}{\varphi_h(\chi_0,\tilde{\theta})} \right)\left( \frac{\chi_0^2}{2d} + \frac{\chi_0}{2d^2} + \frac{1}{4d^3} \right)$$

The value $\chi_0$ should be selected from the conditions $\chi_0^2 >> k_{mn}^2/\sin^2(\theta_0)$, $\chi_0^2 >> k_{pm}^2/\sin^2(\theta_0)$ $k_{pm}^2$. The values $k_{pn}^2$ are related to the concentration of atoms, and the wavelengths $\lambda_{pn} = 2\pi/k_{pn}$ usually correspond to the UV range. The transition frequencies may be higher and correspond to energies of the order of several EV. Therefore, the minimum wavelengths $\lambda_{\min}$ are of the order of several tens of nm, and the magnitude $\chi_0 > 2\pi/\lambda_{\min}$ is of the order of 0.1 (1/nm). This upper limit makes it possible to calculate integrals very accurately.

Consider the behavior of the force at large distances $d$. Making the substitution $\kappa = x/d$, $k = y/d$, we bring (4) to the form

$$P(d) = -\frac{\hbar c}{2\pi^2 d^4} \int_0^\infty x dx \int_0^\infty \sqrt{x^2+y^2} \left( \frac{1}{f_e(x,y,d)} + \frac{1}{f_h(x,y,d)} \right) dy . \qquad (16)$$

For large $d$, the function

$$f_h(x,y,d) = \exp\left(2\sqrt{x^2+y^2}\left[ \frac{2\sqrt{x^2+y^2}\sqrt{x^2+\varepsilon(x,y)y^2} + 2x^2 + (\varepsilon(x,y)+1)y^2}{\varepsilon(x,y)y^2 - y^2} \right]\right)^2 - 1$$

does not depend on this distance. The function $f_e(x,y,d)$ is also independent, so we have $P(d) \sim 1/d^4$. Exponentially small additions of type (15) provide corrections to this dependence. By making the substitution $u = 2K_0 d = 2d\sqrt{\kappa^2 + v^2/d^2}$, $u du/(2d)^2 = \kappa d\kappa$, $v = kd$, we obtain the integrals

$$P(d) = -\frac{\hbar c}{16\pi^2 d^4} \int_0^\infty \int_{2v}^\infty \left( \frac{\varphi_e^{-1}(u,v,d)}{1-\varphi_e^{-1}(u,v,d)\exp(-u)} + \frac{\varphi_h^{-1}(u,v,d)}{1-\varphi_h^{-1}(u,v,d)\exp(-u)} \right)\exp(-u)u^2 dv du .$$

With a large distance $d$, the functions $\varphi_{e,h}$ cease to depend on it, and the integral over $u$ can be approximately calculated by integrating parts three times and discarding the small remainder. Denoting the parenthesis as $\Phi(u,v)$, we get

$$P(d) \approx -\frac{\hbar c}{16\pi^2 d^4} \int_0^\infty dv \exp(-2v)\left[ v^2\Phi(2v,v) + \delta_u\left(u^2\Phi(u,v)\right)_{u=2v} + \delta_u^2\left(u^2\Phi(u,v)\right)_{u=2v} \right].$$

The first term in the square bracket has a second-order zero at zero, so the integral of it can also be approximately calculated by integrating parts three times. As a result, we have a nonintegrative term $\Phi(0,0)/4$ and a contribution to the integral from $\left[4v\Phi'(2v,v)+v^2\Phi''(2v,v)\right]/8$. A stroke means differentiation by the first variable. The second and third terms are equal to $2v\Phi(2v,v)+v^2\Phi'(2v,v)$ and $2\Phi(2v,v)+4v\Phi'(2v,v)+v^2\Phi''(2v,v)$. They also have first- and second-order zeros, so the process can be continued. As a result, it is possible to obtain the decomposition of the derivatives of the function $\Phi$ at zero. If $\varepsilon \to \infty$, then $\varphi_{e,h} \to 1$, and after substitution $K_0 = pk$, $\kappa = k\sqrt{p^2-1}$ we have the Casimir result $P(d) = -\hbar c \pi^2 /\left(240 d^4\right)$.

Consider the following from (4) the Lifshitz problem:

$$P(d) = -\frac{\hbar c}{2\pi^2}\int\limits_1^\infty p^2 dp \int\limits_0^\infty k^3 \left( \frac{1}{\exp(2pkd)S^2(p,1)-1} + \frac{1}{\exp(2pkd)S^2(p,\varepsilon)-1} \right) dk,$$

$s(p) = \sqrt{p^2 - 1 + \varepsilon}$, $S(p,\varepsilon) = (s(p) + \varepsilon(k)p)/(s(p) - \varepsilon(k)p)$. Making a substitution $k = v/d$, we get

$$P(d) = -\frac{\hbar c}{2\pi^2 d^4}\int\limits_1^\infty p^2 dp \int\limits_0^\infty v^3 \left( \left[\exp(2pv)S^2(p,1)-1\right]^{-1} + \left[\exp(2pv)S^2(p,\varepsilon)-1\right]^{-1} \right) dv.$$

Assuming that the main contribution takes place at $p \approx 1$ and counting $s(1) = \sqrt{\varepsilon}$, we have

$$P(d) \approx -\frac{\hbar c}{2\pi^2 d^4}\int\limits_1^\infty p^2 dp \int\limits_0^\infty v^3 \left( \left[\exp(2pv)S^2(1,1)-1\right]^{-1} + \left[\exp(2pv)S^2(p,\varepsilon)\right]^{-1} \right) dv,$$

$S(1,\varepsilon) = \left(\sqrt{\varepsilon(v)} + \varepsilon(v)\right)/\left(\sqrt{\varepsilon(v)} - \varepsilon(v)\right)$. Ignoring the units, we find

$$P(d) \approx -\frac{\hbar c}{2\pi^2 d^4}\int\limits_1^\infty p^2 dp \int\limits_0^\infty v^3 \times \left( \exp(-2pv)\left(\frac{\sqrt{\varepsilon(v)}-1}{\sqrt{\varepsilon(v)}+1}\right)^2 + \right.$$
$$\left. + \exp(-2pv)\left(\frac{\sqrt{\varepsilon(v)}-\varepsilon(v)}{\sqrt{\varepsilon(v)}+\varepsilon(v)}\right)^2 \right) dv .$$

Calculating the integrals with respect to p, we obtain

$$\int\limits_1^\infty p^2 \exp(-2pv)dp = \exp(-2v)\left[\frac{1}{2v} + \frac{1}{2v^2} + \frac{1}{4v^3}\right].$$

The result can be easily obtained if a low-frequency DP is used in the entire range where dissipation occurs $\varepsilon(v) \approx \varepsilon(0)$:

$$P(d) \approx -\frac{3\hbar c}{8\pi^2 d^4}\left(\frac{1-\sqrt{\varepsilon(0)}}{1+\sqrt{\varepsilon(0)}}\right)^2. \qquad (17)$$

This requires that the value $v = \xi d / c$ be small, i.e. $d < c / \xi_{max}$. If the transition frequencies lie in the UV region and are on the order of $10^{16}$ Hz, this means distances of $d < 30$ nm. For diamond $\varepsilon(0) = 5.6$, we obtain a force 0.448 times less than in the Casimir model. Assuming as in [16] $s(p) = p$, we find after the substitution $2pv = x$

$$P(d) \approx -\frac{\hbar c}{16\pi^2 d^3}\int\limits_0^\infty \int\limits_{\xi d/c}^\infty \frac{x^2}{\exp\left(x\left(\frac{1+\varepsilon(v)}{1-\varepsilon(v)}\right)^2\right)-1}\,dvdx.$$

There is a lower limit in this formula $v = \xi d / c$, so it coincides with the formula from [16] (in the latter, the lower limit is taken as zero), i.e. it gives a dependence $1/d^3$. However, this is a transitional dependence from large to small distances. For very small $d$, the force is finite. Also assuming $\varepsilon(v) \approx \varepsilon(0)$ we find

$$P(d) \approx -\frac{\hbar}{16\pi^2 d^3}\left(\frac{1-\varepsilon(0)}{1+\varepsilon(0)}\right)^2 \int\limits_0^\infty \int\limits_{2\xi d/c}^\infty \frac{x^2\exp(-x)}{1-\exp\left(-x\left(\frac{1-\varepsilon(0)}{1+\varepsilon(0)}\right)^2\right)}\,d\xi dx\,,$$

$$P(d) \approx -\frac{\hbar}{16\pi^2 d^3}\left(\frac{1-\varepsilon(0)}{1+\varepsilon(0)}\right)^2 \int\limits_0^\infty d\xi \int\limits_{2\xi d/c}^\infty x^2\left(\exp(-x)+\exp(-2x)\left(\frac{1-\varepsilon(0)}{1+\varepsilon(0)}\right)^2\right)dx\,.$$

The integral over $x$ has the value

$$\int\limits_{2\xi d/c}^\infty x^2\left(\exp(-x)+\exp(-2x)\left(\frac{1-\varepsilon(0)}{1+\varepsilon(0)}\right)^2\right)dx =$$
$$= \exp(-2\xi d/c)\left[(2\xi d/c)^2 + 2(2\xi d/c)+2\right] + \qquad .$$
$$+ \exp(-4\xi d/c)\left[\frac{(2\xi d/c)^2}{2}+\frac{(2\xi d/c)}{2}+\frac{1}{4}\right]\left(\frac{1-\varepsilon(0)}{1+\varepsilon(0)}\right)^2$$

Now the integral over $\xi$ is also easily calculated , which gives terms proportional to $(c/d)^\nu$, $\nu = 0,1,2$, i.e., in addition to the term $1/d^3$, there are terms with $1/d^4$ and $1/d^5$ by taking into account the lower limit. We do not provide the final result.

Теперь легко вычисляется и интеграл по $\xi$, который дает члены, пропорциональные

The Van Kampen method with functions of type (2), (3) (respectively, and the Lifshitz formula) formally does not allow calculating the result in the limit of small $d$. Indeed, it is based on the principle (or theorem) of the argument and requires the vanishing of the integral on the large right semicircle of the complex plane $\xi$ (or $k$). This provides a large multiplier $\exp(2K_0 d) = \exp\left(2\sqrt{\kappa^2 + k^2}\, d\right)$ in the denominator. However, when $d = 0$ it is equal to one and does not ensure convergence. Accordingly, it cannot be decomposed in $d$, and for small $d$, the upper limit should be increased from the condition $\chi_0 > 2\pi / d_{\min}$. So, for $d = 1$ nm we have $\chi_0 > 2\pi$ (1/nm). Since at high frequencies in the model (12) $\varepsilon \to 1$, $\varepsilon(k) - 1 \approx k_{\max}^2 / k^2$, then $K = K_0\left(1 + k_{\max}^2 / \left(2k^2 K_0\right)\right) \to K_0$, $K_0 \to \infty$, and when $d << \sqrt{\kappa_{\max}^2 + k_{\max}^2} / 2$ we obtain

$$\frac{1}{f_h(\kappa, k, 0)} = \frac{k_{\max}^4}{16\left(\kappa^2 + k^2\right)^2 - k_{\max}^4} \to 0,$$

$$\frac{1}{f_e(\kappa, k, 0)} = \frac{\left(2\kappa^2 + k^2\right)^2 k_{\max}^4}{4K_0^4 k^4 - \left(2\kappa^2 + k^2\right)^2 k_{\max}^4} \to 0.$$

However, at $d = 0$, the result (4) does not exist. Indeed, the remainder of the integral from $K_0 f_h^{-1}(\kappa, k, 0)$ at high frequencies is

$$\frac{k_{\max}^4}{16} \int_0^\infty \kappa d\kappa \int_{k_{\max}}^\infty \frac{K_0 dk}{K_0^4 - k_{\max}^4 / 16} \approx \frac{k_{\max}^4}{16} \int_0^\infty d\kappa \left(\frac{1}{\kappa} - \frac{k_{\max}}{\kappa \sqrt{\kappa^2 + k_{\max}^2}}\right).$$

It is logarithmically diverging. The remainder for $f_e^{-1}(\kappa, k, 0)$ also diverges. In principle, integrals can be calculated for any small but finite $d$. But as $d$ decreases, the upper limit should be increased proportionally $1/d$.

Consider the case of a dielectric with DP $\tilde{\varepsilon}(\omega)$ between the plates. In this case, instead $K_0$, we should take in exponent $\tilde{K} = \sqrt{\kappa^2 + \tilde{\varepsilon} k^2}$ and functions of the form [13]

$$f_h(\kappa, k, d) = \exp\left(2\tilde{K} d\right)\left[\frac{K\left(\tilde{K} + K_0\right)\coth(Kt) + K^2 + \tilde{K}K_0}{K\left(\tilde{K} - K_0\right)\coth(Kt) + K^2 - \tilde{K}K_0}\right]^2 - 1, \qquad (18)$$

$$f_e(\kappa, k, d) = \exp\left(2\tilde{K} d\right)\left[\frac{\varepsilon K\left(\tilde{K} + \tilde{\varepsilon}K_0\right)\coth(Kt) + \tilde{\varepsilon}\tilde{K}K_0 + \varepsilon^2 K^2}{\varepsilon K\left(\tilde{K} - \tilde{\varepsilon}K_0\right)\coth(Kt) + \tilde{\varepsilon}\tilde{K}K_0 - \varepsilon^2 K^2}\right]^2 - 1. \qquad (19)$$

For thick layers ($\coth(Kt) = 1$) and a thin film of thickness $d$ with DP $\tilde{\varepsilon}$ between them, we obtain the result [11] corresponding to the result of [21]:

$$P(d) \approx -\frac{\hbar}{16\pi^2 d^3} \int_0^\infty d\xi \int_{2\tilde{\xi}d/c}^\infty \left\{\exp(x)\left(\frac{[\varepsilon(\xi) + \tilde{\varepsilon}(\xi)]^2}{[\varepsilon(\xi) - \tilde{\varepsilon}(\xi)]^2}\right) - 1\right\} x^2 dx.$$

In [11], the result is given for different DP halfspaces, but with the loss of a multiplier of 2. The absence of a film $\tilde{\varepsilon} = 1$ corresponds to the Lifshitz result (3.1) at $\varepsilon_1 = \varepsilon_2 = \varepsilon$ and the zero lower limit. In the case $t = 0$ from (18), (19) we obtain

$$f_h(\kappa, k, d) = \exp\left(2\tilde{K} d\right)\frac{\left(\tilde{K} + K_0\right)^2}{\left(\tilde{K} - K_0\right)^2} - 1,$$

$$f_e(\kappa, k, d) = \exp(2\tilde{K}d)\frac{(\tilde{K} + \tilde{\varepsilon}K_0)^2}{(\tilde{K} - \tilde{\varepsilon}K_0)^2} - 1.$$

The result (4) with these functions corresponds to the external Casimir pressure on a film of thickness $d$ with DP $\tilde{\varepsilon}$ located in a vacuum. For a film with a very large DP $\tilde{\varepsilon} = 1 + k_{max}^2/k^2$ (for a dense plasma), we have $f_{e,h}(\kappa, k, d) \approx \exp\left(2kd\sqrt{p^2 + k_{max}^2/k^2}\right) - 1$ (at $k < k_{max}$). Making a substitution $y = 2kd\sqrt{p^2 + \tilde{\varepsilon}(k) - 1}$, or $y^2 = 4d^2(k^2p^2 + k_{max}^2)$, we get

$$P(d) = -\frac{\hbar c}{16\pi^2 d^4}\int_1^\infty p^{-2}dp \int_{2dk_{max}}^\infty \frac{y^2\sqrt{y^2 - 4d^2k_{max}^2}}{\exp(y) - 1}\,dy.$$

For a thin layer $dk_{max} \to 0$, we get the pressure $P(d) = -\hbar c\pi^2/(1920d^4)$. In the case of a large number of layers, the characteristic equation is obtained by the transmission matrix method [10, 14]. For the Lifshitz problem, it is easier to obtain the characteristic equation by transforming the impedance. So, the normalized E-mode impedance is $\rho_e = \sqrt{\kappa^2 + k^2\varepsilon}/(k\varepsilon)$, and the H-mod impedance is $\rho_h = k/K$. For an empty space (slot) $\rho_{0e} = K_0/k$, $\rho_{0h} = k/K_0$. The impedances $\rho_{e,h}$ are transformed by the slot to the impedance

$$Z = \rho_{0e,h}\frac{\rho_{e,h} + i\rho_{0e,h}\tan(k_{z0}d)}{\rho_{0e,h} + i\rho_{e,h}\tan(k_{z0}d)} =$$
$$= \rho_{0e,h}\frac{\rho_{e,h} + \rho_{0e,h}\tanh(K_0d)}{\rho_{0e,h} + \rho_{e,h}\tanh(K_0d)}\,.$$

Here $k_{z0} = -iK_0$. For resonance, it is necessary $Z = -\rho_{e,h}$, from where we get the equation

$$\tilde{f}_{e,h}(\kappa, k, d) = \rho_{0e,h}\frac{\rho_{e,h} + \rho_{0e,h}\tanh(K_0d)}{\rho_{0e,h} + \rho_{e,h}\tanh(K_0d)} + \rho_{e,h} = 0\,. \tag{20}$$

We have $\tilde{f}_{e,h}(\kappa, k, \infty) = \rho_{e,h} + \rho_{0e,h}$, $\tilde{f}_{e,h}(\kappa, k, 0) = 2\rho_{e,h}$. However, these functions correspond to functions (6) and (7) up to multipliers. Replacing the hyperbolic tangent by $(\exp(2K_0d) - 1)/(\exp(2K_0d) + 1)$, we find

$$\tilde{f}_{e,h}(\kappa, k, d) = \frac{(\rho_{0e,h} + \rho_{e,h})^2\exp(2K_0d) - (\rho_{0e,h} - \rho_{e,h})^2}{(\rho_{e,h} + \rho_{0e,h})\exp(2K_0d) + \rho_{0e,h} - \rho_{e,h}}\,. \tag{21}$$

Integral (4) with function (21) diverges for any finite or even infinite $d$. According to the principle of the argument, it is determined with precision to a certain value associated with the infinite vacuum energy [12]. The value as the difference

$$\frac{1}{\bar{f}_{e,h}(\kappa,k,d)} = \frac{1}{\tilde{\bar{f}}_{e,h}(\kappa,k,d)} - \frac{1}{\tilde{\bar{f}}_{e,h}(\kappa,k,\infty)} =$$
$$= \frac{2\rho_{0e,h}(\rho_{0e,h} - \rho_{e,h})}{(\rho_{0e,h} + \rho_{e,h})^2 \exp(2K_0 d) - (\rho_{0e,h} - \rho_{e,h})^2} \tag{22}$$

at large $d$ vanishes, i.e. integral (4) exists with it at such distances and describes (up to a factor) the force at large distances. To match the Lifshitz problem $\tilde{\tilde{f}}_{e,h} = f_{e,h}$, we should take the function $\tilde{\tilde{f}}_{e,h}(\kappa,k,d) = 2(\rho_{0e,h} - \rho_{e,h})\bar{f}_{e,h}(\kappa,k,d)/\rho_{0e,h}$. Assuming $\varepsilon \to \infty$ ($\rho_{e,h} \to 0$), we find the correspondence of this function to the Casimir problem: $\tilde{\tilde{f}}_{e,h}(\kappa,k,d) = \exp(2K_0 d) - 1$. In particular, for the Lifshitz problem with zero gap, we obtain

$$f_{e,h} = \frac{(\rho_{0e,h} + \rho_{e,h})^2 - (\rho_{0e,h} - \rho_{e,h})^2}{(\rho_{0e,h} - \rho_{e,h})^2}.$$

At high frequencies $\varepsilon(k) = 1 + k_{max}^2/k^2$, and we have $K \approx K_0 + k_{max}^2/(2K_0)$, where $K_0$ is a large value. Therefore $f_e \approx 4k^2(1 + k_{max}^2/k^2)(K_0^2 + k_{max}^2)/k_{max}^4 \approx 4k^2 K_0^2/k_{max}^4$, and similarly $f_h \approx 16K_0^3(K_0 + k_{max}^2/(2K_0))/k_{max}^4 \approx 16K_0^4/k_{max}^4$. These values are large, but they do not ensure convergence of the integrals. Indeed, consider the integral for H-modes:

$$\int_0^\infty \kappa d\kappa \int_{k \gg k_{max}}^\infty \frac{K_0}{f_h(\kappa,k)} dk \approx k_{max}^4 \int_0^\infty \kappa d\kappa \int_{k \gg k_{max}}^\infty \frac{1}{16K_0^3}\left(1 - \frac{2k_{max}^2}{4K_0^2}\right) dk.$$

Replacing the parenthesis with one, we get a logarithmically divergent integral

$$\frac{k_{max}^2}{16} \int_{k \gg k_{max}}^\infty \frac{dk}{k}.$$

Соответственно метод Ван Кампена не позволяет вычислить силу при бесконечно малом (нулевом) зазоре. Рассмотрим соответствующую функцию (6) и уравнение $f_h(\kappa,k_0) = 0$, принимающее вид $k = \ln(\pm\gamma(k,p))/(pd)$, поскольку

$$\exp(K_0 d) = \exp(kpd) = \pm\left(\frac{K - K_0}{K + K_0}\right) = \pm\gamma(p,k).$$

All branches of the logarithm should be taken into account here. It follows that for small $d$, the zeros are shifted to the high frequency range. The motion of zeros in the complex plane is shown in [12] (Fig. 3). This applies to low frequencies and short distances. For high frequencies $K \approx K_0$, $\gamma(p,k) = 0$, and the value $k = \ln(\pm\gamma)/(pd)$ becomes indeterminate at $d \to 0$. At low $d$, all frequencies become large, and the plates are transparent to them. This suggests that the value $P(0)$ is finite. Indeed, the infinite attraction of the two plates would release infinite energy, which is physically absurd. Although, on the other hand, the continuum model no longer holds in this case. The frequencies can be found by solving the equation ( $n = 0, \pm 1 \pm \dots$ )

$$k = \frac{i\omega}{c} = \frac{1}{pd} \ln\left( \pm \frac{\sqrt{p^2 - 1 + \varepsilon(k)} - p}{\sqrt{p^2 - 1 + \varepsilon(k)} + p} \right) + \frac{2i\pi n}{pd}.$$

Hence , for the real parts we have $\omega_n' = \pm 2\pi n/(pd)$, $n = 1,2,\dots$, and for the imaginary parts

$$\omega_n'' = \pm \frac{1}{pd} \ln\left( \frac{\sqrt{p^2 - 1 + \varepsilon(k)} - p}{\sqrt{p^2 - 1 + \varepsilon(k)} + p} \right).$$

Thus, at small distances, all frequencies are shifted to an infinite region. Infinite frequencies are not perturbed by the dielectric, so the contribution to the perturbation energy is zero or at least finite. On the other hand, equation (20) $\tilde{f}_{e,h}(\kappa,k,0) = 0$ implies $\rho_{e,h} = 0$, or $\rho_e = \sqrt{p^2 - 1 + \varepsilon(k)} = 0$. In this case, at $\varepsilon(k) > 1$ the equation has no zeros in the finite domain. The equation $\rho_h = 0$ has zero at $p = \infty$, Similarly, for two plates at $d = 0$, all resonant frequencies are shifted to infinity, so the force density is finite.

For the Lifshitz problem, at $K_0 > k_{max}$ we have

$$\frac{1}{f_h(\kappa,k)} \approx \frac{k_{max}^4/\left(4K_0^2\right)}{\exp(2K_0 d)\left(4K_0^2 + 2k_{max}^2\right)^2 - k_{max}^4/\left(4K_0^2\right)} \approx$$
$$\approx \frac{k_{max}^4}{16K_0^4 \exp(2K_0 d)}\left(1 + \frac{k_{max}^4}{16K_0^4 \exp(2K_0 d)}\right).$$

Convergence will be if $\exp(2K_0 d) \geq (2K_0 d)^\nu$. We have the equation $\exp(x) = x^\nu$ and its root $x_0 = 2K_0 d$, $x_0 \approx 2$, for $\nu = 3$. We obtain the convergence condition of the integral for small $d$: $\kappa^2 + k^2 >> x_0^2/4d^2 \approx 1/d^2$. It is the same for $f_e(\kappa,k_0)$.

In the case of a thermal field (thermostat) with temperature $T$, the average energy of the quantum oscillator has the form [3]

$$\Theta(\tilde{\omega}_n) = \hbar\tilde{\omega}_n\left[ \frac{1}{2} + \frac{1}{1 + \exp(\hbar\tilde{\omega}_n/(k_B T)) - 1} \right] =$$
$$= \frac{\hbar\tilde{\omega}_n}{2} \coth(\hbar\tilde{\omega}_n/(2k_B T))$$
(23)

therefore, in (1), instead of $\hbar\tilde{\omega}_n/2$ should be used $\Theta(\tilde{\omega}_n)$. Note that energy (23) is an even function of frequency (positive for negative frequencies). Also, the functions $f_{e,h}(p,\omega)$ are even. Now (4) should be taken as

$$P(d,T) = -\frac{\hbar}{2\pi^2 c^3} \int_1^\infty p^2 dp \int \omega^3 \coth\left( \frac{\hbar\omega}{2k_B T} \right)\left( \frac{1}{f_e(p,\omega)} + \frac{1}{f_h(p,\omega)} \right)d\omega,$$

or inn the form

$$P(d,T) = -\frac{\hbar c}{2\pi^2} \int\limits_1^\infty p^2 dp \int\limits_0^\infty \cot\left(\frac{\hbar k c}{2 k_B T}\right)\left(\frac{1}{f_e(p,k)} + \frac{1}{f_h(p,k)}\right) k^3 dk , \qquad (24)$$

which coincides with the Lifshitz formula for the final temperature [16]. In the case of high temperatures $k_B T \gg \hbar\omega$, the frequencies under consideration (hot plasma) will be

$$P(d,T) = -\frac{k_B T}{\pi^2} \int\limits_1^\infty p^2 dp \int\limits_0^\infty \left(\frac{1}{f_e(p,k)} + \frac{1}{f_h(p,k)}\right) k^2 dk .$$

We find a correction to formula (4) at a small finite (on the order of room) temperature using decomposition $\cot(x) \approx (1 + \exp(-2x))^2 \approx 1 + 2\exp(-2x)$ at large $x$. We have

$$P(d,T) = P(d,0) - \frac{\hbar}{\pi^2 c^3} \int\limits_1^\infty p^2 dp \int \omega^3 \exp\left(-\frac{\hbar\omega}{k_B T}\right)\left(\frac{1}{f_e(p,\omega)} + \frac{1}{f_h(p,\omega)}\right) d\omega .$$

Integral (24) has poles. $k_n = 2n\pi k_B T /(\hbar c) = i\omega_n / c$ It can be calculated by replacing, as usual, the integral by the sum of the Matsubara frequencies $\omega_n = i\xi_n = ick_n$ [15,16], i.e. by taking half-residues at $\omega_n$ and a quarter of residue at $\omega_0 = 0$ :

$$P(d) = -\frac{8\pi^2 (k_B T)^4}{c^3 \hbar^3} \sum_{n=0}^\infty \frac{1}{1 + \delta_{n0}} \int\limits_1^\infty \left(\frac{1}{f_e(x/n,\omega_n)} + \frac{1}{f_h(x/n,\omega_n)}\right) x^2 dx . \qquad (25)$$

Here in $f_{e,h}$ the values $K(x/n,\omega_n) = k_n \sqrt{x^2 / n^2 + \varepsilon - 1}$ are included. Since the integral function in (24) is even in $k$, the integral can be extended to the entire axis and the integration contour can be taken as shown in Fig. 3.7 of [15]. Since the frequencies $\omega_{mnl} = \omega'_{mnl} \pm i\omega''_{mnl}$ are located in the right half-plane $\omega$ and are complex conjugate, such integration yields a doubled sum over the positive frequencies. Therefore, the real parts in Van Kampen's formulas do not need to be taken, and the integral functions are real. Since the poles are simple, enclosing them with small neighborhoods, integral (24) can be calculated numerically in the sense of the main value. In formula (25), n=0 corresponds to $f_e(\infty,\omega_n) = \infty$, and this contribution, as it is easy to see, is absent. Thus, the Van Kampen method allows considering an arbitrary number of layers by constructing a characteristic equation, as well as taking temperature into account. It also allows you to insert conductive, for example, graphene sheets into the layers. In the simplest case of weighted sheets with normalized conductivity $\zeta$, we have

$$f_{e,h} = \frac{(\varsigma\rho_{e,h})^2}{(2 + \varsigma\rho_{e,h})^2 \exp(2K_0 d) - (\varsigma\rho_{e,h})^2} .$$

The summation results [11] correspond to the structures of Casimir, Lifshitz, and others.

There is a very extensive literature on the Casimir effect, which is difficult to cover, including monographs and reviews, for example, [12,15,22–30]. The problem of complex frequencies in dissipative structures is widely discussed in it, but little attention is paid to the method of work [11]. Meanwhile, it allows obtaining correct results in dissipative structures [11, 12], and using an arbitrary number of layers. This is because the principle of the argument gives a real sum of complex conjugate frequencies. In reality, there are both frequencies in DE. If $\tilde{\omega}_n = \tilde{\omega}'_n + i\tilde{\omega}''_n$ are responsible for absorption (time-damping waves) by atoms, then $\tilde{\omega}_n^* = \tilde{\omega}'_n - i\tilde{\omega}''_n$ correspond to radiation (increasing waves). According to Kirchhoff's law, at thermodynamic equilibrium, radiation at each frequency is exactly equal to absorption, so $\hbar \text{Re}(\tilde{\omega}_n)/2$ is exactly equal to the stored energy, just as $\varepsilon_0 \text{Re}(\varepsilon |\mathbf{E}(\omega)|^2)/2$ there is an average stored field energy over the period [4–9] in media where there is no energy accumulation due to particle motion (in this case, always $\text{Re}(\varepsilon) > 1$, unlike plasma, where there may be $\text{Re}(\varepsilon) < 0$ or for a Lorentz oscillator in a narrow range near resonance). A number of papers have proved the Lifshitz formula, for example, in [3]. In particular, in the works of Schwinger [31–34], an approach based on field sources was used. In [35], a general theory based on the Green's functions of the quantum field statistical approach was developed. In [11,12,36–39], it was proved on the basis of the principle of the argument. In [40], a rather complex approach was proposed for the quantum interaction of a damped oscillator with the thermal field of a thermostat using the Zwanzig–Caldeira–Leggett quantum model. The Van Kampen method makes it possible to circumvent the problem of absorption when calculating Casimir forces. Note that in the Lifshitz formula itself, the real part is taken, but when moving to the plane of the imaginary frequency, the result becomes valid.

The numerical results are shown in Figures 1 and 2. The following DP(9) model is used: $m = 1$, $n = 6$, $k_p = 0.05$, $k_{pn} = 0.05$, $n = 1,...,6$, $k_{r1} = 0.01$, $k_{r2} = 0.02$, $k_{r3} = 0.03$, $k_{r4} = 0.04$, $k_{r5} = 0.05$, $k_{r6} = 0.08$, $k_{c0} = k_{cn} = 10^{-6}$, (everything is in reverse nm). The dependence of pressure $P$ on the thickness of the plates at different distances is shown in Fig. 1. Curves 1–4 correspond to the absence of conductivity: $k_p = 0$, $k_{pn} = 0.05$, curve 5 is constructed at $k_p = 0.05$. All curves are saturated at thicknesses of the order of 10 nm, so measurements with such plates or thicker give the force in Lifshitz configuration force. The dependence of pressure $P$ on distance at different plate thicknesses is shown in Fig. 2. One can see the difference at short distances from the law $1/d^4$, which is carried out at long distances. This difference is already strongly evident at $d < 10$ nm. When $d \to 0$, the pressure tends to the finished value.

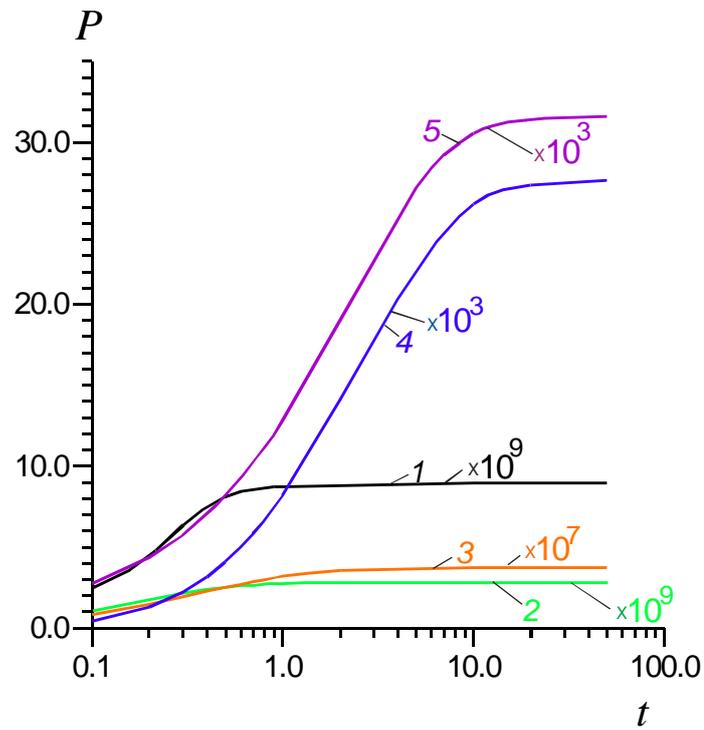

Fig. 2. Casimir pressure $P=F/L^2$ (N/m$^2$) between two dielectric layers depending on their thickness $t$ (nm) at different distances $d$ (nm): $d$=0.01 (curve 1); 0.1 (2); 1 (3); 10 (4,5). Curves 1–4 are plotted in the absence of conductivity ($\omega_p = 0$), curve 5 – in the presence of $\omega_p$

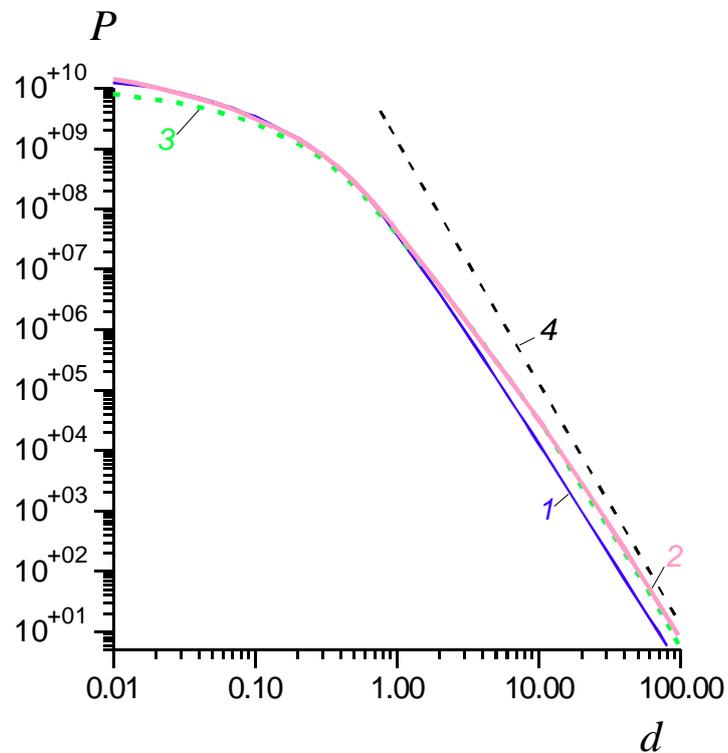

Fig. 3. Casimir pressure $P$ (N/m$^2$) between two dielectric layers depending on the distance $d$ (nm) at different thicknesses $t$ (nm): $t$=1 (curve 1); 50 (2,3). Curves 1, 2 are constructed taking into account the conductivity, curve 3 at . 4 – the Casimir result

The case of the absence of conductivity $k_p = 0$ is also considered there (curve 3). At nm $t \sim 1$ or less, the results for the van der Waals force are completely obtained by the method of density functional theory and correspond to the above. Integrals of type (4) were calculated by replacing $\kappa = \chi \cos(\theta)$, $k = \chi \sin(\theta)$ using 600 points of integration along the angle and 5000 points of integration along $\chi$, and the region was divided into 6 subdomains with simultaneous integration into them. The lower area matched $0 < \chi < k_{c0}$. The upper area corresponded to $k_{r6} < \chi < k_{\max}$, where $k_{\max} = 1 + 10 k_{r6} + 1/d$. The choice of the specified number of points guaranteed an accuracy of three decimal places.

**Financing the work**

The work was carried out with the financial support of the Ministry of Education and Science of the Russian Federation within the framework of the state assignment (FSRR-2023-0008).

———————————

1. H.B.G. Casimir, On the attraction between two perfectly conducting plates, Proc. K. Ned. Akad. Wet. **51**, 793–795 (1948).

2. H.B.G. Casimir, D. Polder, The Influence of Retardation on the London-van der Waals Forces, Phys. Rev. E **73**(4), 360–372 (1948). DOI: 10.1103/PhysRev.73.360.

3. M.V. Levin, S.M. Rytov, *Theory of equilibrium thermal fluctuations in electrodynamics* (Nauka, Moscow, 1967). [In Russian].

4. A.I. Akhiezer I.A. Akhiezer, *Electromagnetism and electromagnetic waves* (Higher School, Moscow, 1985). [In Russian].

5. L.D. Goldstein, N.V. Zernov, Electromagnetic fields and waves (Soviet Radio, Moscow, 1971). [In Russian].

6. L.A. Vainstein, Electromagnetic waves (Radio and Communications, Moscow, 1988). [In Russian].

7. M. V. Davidovich, Electromagnetic Energy Density and Velocity in a Medium with Anomalous Positive Dispersion, Technical Physics Letters **32**(11), 982–986 (2006). DOI: 10.1134/S106378500611023X.

8. M. V. Davidovich, On the Electromagnetic Energy Density and Energy Transfer Rate in a Medium with Dispersion due to Conduction, Technical Physics 55( 5), 630–635 (2010). DOI: 10.1134/S1063784210050063.

9. M. V. Davidovich, On Times and Speeds of Time-Dependent Quantum and Electromagnetic Tunneling, Journal of Experimental and Theoretical Physics 130(1), 35–51 (2020). DOI: 10.1134/S1063776119120161.

10. A.L. Feldstein, L.R. Yavich. *Synthesis of four-pole and eight-pole microwave devices* (Svyaz, Moscow, 1971). [In Russian].

11. N.G. Van Kampen, B.R.A. Nijboer, K. Schram, On the macroscopic theory of van der Waals forces, Phys. Lett. A **26**, 307–308 (1968). DOI: 10.1016/0375-9601(68)90665-8.

12. S.K. Lamoreaux, The Casimir force: Background, experiments and applications, Reps. Progr. Phys. **65**, 201–236, (2005). DOI:10.1088/0034-4885/68/1/R04.

13. V.V. Bryksin, M. Petrov, Casimir force with the inclusion of a finite thickness of interacting plates, Physics of the Solid State **50**(2), 229–234 (2008). DOI: 10.1134/S1063783408020029.

14. M.V. Davidovich, Plasmons in multilayered plane-stratified structures, 2017 Quantum Electron. **47**(6), 567–579 (2017). DOI: 10.1070/QEL16272.

15. W. Simpson, U. Leonhardt, *Forces of the quantum vacuum: an introduction to Casimir physics* (The Weizmann Institute of Science, Israel, 1965).

16. E.M. Lifshitz, The Theory of Molecular Attractive Forces between Solids, Sov. Phys. JETP, Vol. 2, 1956, pp. 73–83. Sov. Phys. JETP 2, 73 (1956).

17. S.V. Gaponenko, D.V. Novitsky, Wigner time for electromagnetic radiation in plasma, Phys. Rev. A 106, 023502 (2022). DOI: 10.1103/PhysRevA.106.023502.

18. T.V. Sobolev, A.P. Timonov, V.Val. Sobolev, Fine Structure of the Dielectric-Function Spectrum in Diamond, Semiconductors **34**(8), pp. 902–907 (2000). DOI: 10.1134/1.1188098.

19. T.L. Cocker, D. Baillie, M. Buruma, L.V. Titova, R.D. Sydora, F. Marsiglio, F.A. Hegmann, Microscopic origin of the Drude-Smith model, Phys. Rev. B **96**, 205439 (2017). DOI: 10.1103/PhysRevB.96.205439.

20. W.-C. Chen, R.A. Marcus, The Drude-Smith Equation and Related Equations for the Frequency-Dependent Electrical Conductivity of Materials: Insight from a Memory Function Formalism, Chemphyschem **22**(16),1667–1674 (2021). DOI: 10.1002/cphc.202100299.

21. I.E. Dzyaloshinskii, E.M. Lifshitz, L.P. Pitaevskii, Van Der Waals Forces in Liquid Films, Soviet Phys. JETP **10**, 161–170 (1960).

22. K.A. Milton, *The Casimir Effect: Physical Manifestations of Zero-Point Energy* (Singapore: World Scientific, 2001).

23. G.L. Klimchitskaya, A.B. Fedortsov, Y.V. Churkin, V.A. Urova, Casimir force pressure on the insulating layer in metal-insulator-semiconductor structures. Phys. Solid State **53**, 1921–1926 (2011). DOI: 10.1134/S1063783411090174.

24. M. Bordag (ed.), *The Casimir Effect 50 Years Later* (World Scientific, Singapore, 1999).

25. B. Geyer, G. L. Klimchitskaya, V. M. Mostepanenko, Thermal quantum field theory and the Casimir interaction between dielectrics, Phys. Rev. D 72, 085009 (2005). DOI: 10.1103/PhysRevD.72.085009.

26. G. Plunien, B. Muller, W. Greiner, The Casimir effect, Phys. Rep. **134**, 87 (1986). DOI:10.1016/0370-1573(86)90020-7.

27. M. Bordag, U. Mohideen, and V. M. Mostepanenko, New developments in the Casimir effect, Phys. Rep. 353, 1 (2001). DOI: 10.1016/S0370-1573(01)00015-1.

28. K.A. Milton, The Casimir effect: Recent controversies and progress, J. Phys. A 37, R209 (2004).

29. M. Bordag, G. Klimchitskaya, U. Mohideen, V. Mostepananko, *Advances in the Casimir effect*, Int. Ser. Monogr. Phys. 145, 1 (2009). https://inspirehep.net/ literature/841474.

30. V.M. Mostepanenko, N.N. Trunov, The Casimir effect and its applications, Sov. Phys. Usp. **31** 965–987 (1988). DOI: 10.1070/PU1988v031n11ABEH005641.

31. J. Schwinger, Casimir effect in source theory, Letters in Mathematical Physics **1**, 43–47 (D. Reidel Publishing Company, 1975).

32. J. Schwinger, L.L. DeRaad, K.A. Milton, Casimir effect in dielectrics, Ann. Phys. **115**(1), 676–698 (1978). DOI: 10.1016/0003-4916(78)90172-0.

33. J. Schwinger, Casimir Effect in Source Theory II, Letters in Mathematical Physics **24**, 59-61 (Kluwer Academic Publishers, 1992).

34. K.A. Milton, Julian Schwinger and the Casimir Effect: The Reality of Zero-Point Energy, arXiv:hep-th/9811054 (1998).

35. I.E. Dzyaloshinskii, E.M. Lifshitz, L.P. Pitaevskii, General theory of van der Waals' forces, Physics–Uspekhi, **4**(2), 153–176 (1961). DOI: https://doi.org/10.1070.

36. P.W. Milonni, *The Quantum Vacuum* (San Diego, CA: Academic, 1994).

37. K. Schram, On the macroscopic theory of retarded Van der Waals forces, Physics Letters A **43**(3), 282–284 (1973). DOI: 10.1016/0375-9601(73)90307-1%20.

38. F. Intravaia, R. Behunin, Casimir effect as a sum over modes in dissipative systems, Phys. Rev. A **86**, 062517 (2012) DOI: 10.1103/PhysRevA.86.062517.

39. F. Intravaia, How modes shape Casimir physics, International Journal of Modern Physics A **37**(19), 2241014 (2022). DOI: 10.1142/S0217751X22410147.

40. Y.S. Barash, Damped Oscillators within the General Theory of Casimir and van der Waals Forces. J. Exp. Theor. Phys. **132**, 663–674 (2021). DOI: 10.1134/S1063776121040014.